\documentclass{nature}

\usepackage{amsmath}
\usepackage{amssymb}
\usepackage{mathtools}
\usepackage{tabularx}
\usepackage{url}
\newcommand{\dif} {\mathrm{d}}

\newcommand{\Pf}[1]{\text{Pf}\left( #1\right) }

\newcommand{\mean}[1]{\left\langle #1 \right\rangle}
\renewcommand{\vec}{\mathbf}

\pdfpageattr{/Group << /S /Transparency /I true /CS /DeviceRGB>>} 

\usepackage{color}
\definecolor{darkgreen}{rgb}{0.1,0.8,0} 
\definecolor{violet}{rgb}{0.5,0,0.5}
\definecolor{orange}{rgb}{0.8,0.5,0.2}
\definecolor{grey}{rgb}{0.7,0.7,0.7}
\usepackage[normalem]{ulem}

\title{Evolutionary games of condensates in coupled birth-death processes}
\author{Johannes Knebel$^{1\dagger}$, Markus F. Weber$^{1\dagger}$, Torben Kr\"uger$^{2\dagger}$ \& Erwin Frey$^{1\star}$}

\begin{document}

\maketitle

\begin{affiliations}
\item Arnold Sommerfeld Center for Theoretical Physics and Center for NanoScience, Department of Physics, Ludwig-Maximilians-Universit\"at M\"unchen, Theresienstra\ss e 37, 80333 M\"unchen, Germany.
\item 
IST Austria, Am Campus 1, 3400 Klosterneuburg, Austria.
\end{affiliations}

\noindent\upshape$^{\dagger}$These authors contributed equally to this work.\\
\noindent\upshape$^{\star}$Corresponding author: frey@lmu.de.

\vspace{12pt}
\noindent \textbf{Final manuscript with high-resolution figures and electronic supplementary material published in: Nature Communications 6:6977 (2015), doi:10.1038/ncomms7977 }\\
\noindent\textbf{\url{www.nature.com/ncomms/2015/150424/ncomms7977/full/ncomms7977.html} \\(open access)}

\clearpage

\begin{abstract}
Condensation phenomena arise through a collective behaviour of particles. They are observed in both classical and quantum systems, ranging from the formation of traffic jams in mass transport models to the macroscopic occupation of the energetic ground state in ultra-cold bosonic gases (Bose-Einstein condensation). 
Recently, it has been shown that a driven and dissipative system of bosons  may form multiple condensates. Which states become the condensates has, however, remained elusive thus far. The dynamics of this condensation are described by coupled birth-death processes, which also occur in evolutionary game theory. 
Here, we apply concepts from evolutionary game theory to explain the formation of multiple condensates in such driven-dissipative bosonic systems. We show that vanishing of relative entropy production determines their selection. The condensation proceeds exponentially fast, but the system never comes to rest. Instead, the occupation numbers of condensates may oscillate, as we demonstrate for a rock-paper-scissors game of condensates.

\end{abstract}

Condensation  phenomena occur in a broad range of contexts in both classical and quantum systems.
Networks such as the World-Wide-Web or the citation network perpetually grow by the addition of nodes or links and they evolve by rewiring. Over time, a finite fraction of the links of a network may be attached to particular nodes. These nodes become hubs and thereby dominate the dynamics of the whole network; they become condensate nodes\cite{Krapivsky2000,Bianconi2001, Evans2005}. 
Condensation also occurs in models for the jamming of traffic\cite{Evans1996, Krug1996, Chowdhury2000, Kaupuzs2005} and in related mass transport models in which particles hop between sites on a lattice\cite{Evans2005, Spitzer1970, Evans2014}. A condensate forms when a finite fraction of all particles aggregates into a cluster that dominates the total particle flow.
Bose-Einstein condensation, on the other hand, is a quintessentially quantum mechanical phenomenon. When an equilibrated, dilute gas of bosonic particles is cooled to a temperature near absolute zero, a finite fraction of bosons may condense into the energetic ground state\cite{Bose1924,Einstein1924,Einstein1925}. Long-range phase coherence builds up and quantum physics becomes manifest on the macroscopic scale\cite{Griffin1995,Anglin2002}.

In both the classical and the quantum mechanical context, condensation occurs when one or multiple states become macroscopically occupied (they become condensates), whereas the other states become depleted\cite{Penrose1956,Mueller2006}. However, the physical origins of condensation in the above examples differ from each other.
Why and how condensation arises in a particular system remains a topic of general interest and vivid research. 

Here, we study condensation in two systems from different fields of research: incoherently driven-dissipative systems of non-interacting bosons and evolutionary games of competing agents. 
As we show below, the physical principle of vanishing entropy production governs the formation of condensates in both of these systems.
The entities that constitute the respective system shall be called particles. They may be quantum or classical particles (bosons or agents). The dynamics of these particles eventually lead to condensation into particular states (quantum states or strategies). 
Before describing the above two systems, we now introduce the mathematical framework of our study.

On an abstract level, we consider a system of $S$ (non-degenerate) states $E_i, i=1,\dots, S$, each of which is occupied by $N_i\geq 0$ indistinguishable particles, see Fig.~\ref{fig:1}a. The configuration of the system at time~$t$ is fully characterized by the occupation numbers $\vec{N} = (N_1, N_2, \dots, N_S)$. This configuration changes continuously in time due to the transition of particles between states. The total number of particles in this coupled birth-death process is conserved ($N=\sum_i N_i$). 
We are interested in the probability $P(\vec{N}, t)$ of finding the system in configuration $\vec{N}$ at time $t$. The temporal evolution of the probability distribution $P(\vec{N}, t)$ is governed by the classical master equation\cite{Gardiner, VanKampen2007}: 
\begin{align}\label{eq:meq}
\partial_t P(\vec{N}, t) = \sum_{\substack{i,j=1\\ j \neq i}}^S \Big(\Gamma_{i\leftarrow j}(N_i-1,N_j+1)&P(\vec{N}-\vec{e}_i+\vec{e}_j, t)  \notag\\
 -\Gamma_{i\leftarrow j}(N_i,N_j)&P(\vec{N}, t)\Big) \ ,
\end{align} 
where $\vec{e}_i\in \mathbb{Z}^S$ denotes the unit vector in direction $i$ (equal to one at index $i$, otherwise zero).
The rate for the transition of particles from state $E_j$ to $E_i$ depends linearly on the number of particles in the departure and in the arrival state: 
\begin{align}\label{eq:rates}
\Gamma_{i \leftarrow j} = r_{ij} (N_i+s_{ij}) N_j\ ,
\end{align} 
with rate constant $r_{ij}\geq 0$ and constants $s_{ij}\geq 0$. 

Condensation in this framework is understood as the macroscopic occupation of one or multiple states\cite{Penrose1956, Mueller2006}: We consider a state $E_i$ as a condensate when the long-time average of the number of particles in this state scales linearly with the system size ($\langle N_i \rangle_{t}\sim\mathcal{O}(N)$ for large $t$). Hence, a condensate harbours a finite fraction of the total number of particles for large systems ($N\gg 1$).
We refer to a state as depleted when its average occupation number scales less than linearly with the system size. Therefore, the fraction of particles in a depleted state vanishes in the limit of large systems.

Depending on the values of the rate constants $r_{ij}$, numerical simulation of equations~\eqref{eq:meq} with rates~\eqref{eq:rates} reveals that all states, multiple states, or only one state become condensates when the particle density $N/S$ is large enough to detect condensation\cite{Vorberg2013}.  
Thus far, various questions about condensation have remained elusive for the coupled birth-death process defined by equation~\eqref{eq:meq}: Which of the states become condensates? How does this selection of condensates proceed? Is it possible to construct systems that condense into a specific set of condensates?

In the following, we answer these questions by illuminating the physical principle that governs the formation of multiple condensates on the leading order timescale. We show that the vanishing of relative entropy production determines the selection of condensates (see equations~\eqref{eq:ALVE} and~\eqref{eq:D} below). We elaborate how condensate selection is determined by the rate constants $r_{ij}$. The condensation proceeds exponentially fast into a dynamic, metastable steady state within which the occupation numbers of condensates may oscillate.
By applying our general results to systems with many states, we show that the interplay between critical properties of such networks of states\cite{Albert2002} and dynamically stable network motifs\cite{Milo2002} determines the selection of condensates.
The results of our analysis apply to any system whose dynamics are described by the coupled birth-death processes~\eqref{eq:meq} with rates~\eqref{eq:rates}. Before proceeding to the mathematical and numerical analysis of condensation in these processes, we now give a brief overview of such systems.

\section*{Results}
\subsection{Non-interacting bosons in driven-dissipative systems.}

The classical master equation~\eqref{eq:meq} has recently been derived by Vorberg et al. in the study of bosonic systems that are dissipative and driven by external sources\cite{Vorberg2013}. 
For a system of non-interacting bosons that is weakly coupled to a reservoir and driven by an external time-periodic force (a so-called Floquet system)\cite{Bluemel1991, Kohler1997, Breuer2000}, 
one can eliminate the reservoir degrees of freedom (Born and Markov approximation)\cite{Grifoni1998, Breuer2006} and the density matrix of the system becomes diagonal (see the Supplement of the work of Vorberg~et~al.\cite{Vorberg2013}).
The effective dynamics of the bosons become incoherent and are captured on a macroscopic level in terms of the coupled birth-death processes~\eqref{eq:meq} with rates $\Gamma_{i \leftarrow j} = r_{ij} (N_i+1) N_j$ (that is all $s_{ij}=1$ in the rates~\eqref{eq:rates}).
These non-equilibrium setups may not only lead the bosons into a single, but also into multiple condensates\cite{Vorberg2013}. 

For the incoherently driven-dissipative systems described above, the state $E_i$ denotes a time-dependent Floquet state\cite{Bluemel1991, Kohler1997, Breuer2000}. The total rate $\Gamma_{i \leftarrow j}$ for the transition of a boson from state $E_j$ to $E_i$ depends linearly on the number of bosons in the departure state ($N_j$) and the arrival state ($N_i+1$). The latter factor stems from the indistinguishability of bosons and reflects their tendency to congregate. Although, we refer to equation~\eqref{eq:meq} as a classical master equation and coherence does not build up, the quantum statistics of bosons is still encoded in the functional form of $\Gamma_{i \leftarrow j}$. The rate constant $r_{ij}$ is determined by microscopic properties of the system and the reservoir. 

Condensation in the above setup is to be distinguished from Bose-Einstein condensation. 
Typically, studies on Bose-Einstein condensation focus on the existence of long-range phase coherence in thermal equilibrium\cite{Bose1924,Einstein1924,Einstein1925, Penrose1956, Griffin1995,Anglin2002}, its kinetic formation\cite{Griffin1995, Gardiner1997, Kagan1997, Bijlsma2000, Gardiner1998, Walser1999, Kocharovsky2000,Anglin2002}, and the fragmentation of a coherent condensate into multiple condensates (for example when the equilibrium ground state is degenerate)\cite{Griffin1995,Mueller2006}. 
In contrast, the classical birth-death processes~\eqref{eq:meq} with rates~\eqref{eq:rates} describe condensation in bosonic systems that are externally driven by a continuing supply of energy, dissipate into the environment, and exhibit decoherence.


Equations of type~\eqref{eq:meq} may also arise in atomic physics and quantum optics and are known as Pauli master equations\cite{Pauli1928,Mandel1995,Gardiner2004}.
They describe how the population of $S$ non-degenerate energy levels changes over time when a system harbours $N$ indistinguishable, non-interacting bosonic atoms. Such changes may occur by interactions with a radiation field that induces transitions between energy levels. A theoretical description of these transitions in terms of a Pauli master equation is appropriate if coherence is negligible.
As in the previous example, the system then approaches a state in which some of the energy levels are macroscopically occupied (condensates) whereas others are depleted.
More generally, whenever a rate constant $r_{ij}$ governs the transition of a single boson from a state $E_j$ to $E_i$, the rates~\eqref{eq:rates} with $s_{ij}=1$ for all $i$ and $j$ apply if $N$ non-interacting bosons are brought into the system\cite{Vorberg2013}.

\subsection{Strategy selection in evolutionary game theory.}
The classical master equation~\eqref{eq:meq} also occurs in evolutionary game theory (EGT). 
Historically, EGT was developed to study evolutionary processes that are driven by selection and mutation\cite{Smith1982, Nowak2004} and seeks to identify optimal strategies for competitive interactions. For example, EGT has been applied in the study of the prominent ``rock-paper-scissors'' game, which was proposed as a facilitator of species coexistence and has inspired both experimental and theoretical research\cite{Sinervo1996, Kerr2002, Reichenbach2007, Weber2014, Szolnoki2014}. 
Furthermore, the ``prisoner's dilemma'' game serves as a paradigmatic model to explore the evolution and maintenance of cooperation\cite{Nowak2004paper, Szolnoki2014b}. 
The interplay between non-linear and stochastic effects underlies the dynamics of such evolutionary games\cite{McKane2005, Traulsen2005, Reichenbach2006, Melbinger2010, Biancalani2014, Rulands2014}.

In EGT, one typically considers a system of $N$ interacting agents (classical particles) who repeatedly play one fixed strategy $E_i$ out of the $S$ possible choices $E_1, E_2, \dots, E_S$. In each succeeding interaction, the defeated agent adopts the strategy of its opponent. Since $N_j$ agents playing strategy $E_j$ can potentially be defeated by one of the $N_i$ agents playing strategy $E_i$, the rate of change is $\Gamma_{i \leftarrow j} = r_{ij} N_i N_j$. If an agent who plays $E_j$ can also spontaneously mutate into an agent who plays $E_i$ (with rate $\mu_{ij}=r_{ij}s_{ij}$), one recovers the classical master equation~\eqref{eq:meq} with rates~\eqref{eq:rates}.

Thus, there exists a correspondence between condensation in incoherently driven-dissipative bosonic systems and strategy selection in evolutionary game theory: the transition of bosons between states can be interpreted in terms of the interaction and mutation of agents employing evolutionary strategies. In effect, the states in an incoherently driven-dissipative setup play an evolutionary game and the winning states form the condensates. 

After having introduced the above examples, we now proceed with the mathematical and numerical analysis of the classical master equation~\eqref{eq:meq}. We show that the dynamics of condensation change on two distinct timescales. At the leading order timescale, the dynamics are described by a set of nonlinearly coupled, ordinary differential equations (see equation~\eqref{eq:ALVE} below), which determine the states that become condensates. We identify these states by applying concepts from EGT. After an exposition of  the  physical principles that underlie the condensation dynamics, implications of our general results for incoherently driven-dissipative systems are discussed.

\subsection{The antisymmetric Lotka-Volterra equation.}\ 
The total number of particles needed for condensation phenomena to occur is large ($N \gg 1$). In order to detect macroscopic occupancies, it is also assumed that the particle density $N/S$ is large. Therefore, one may approximate the classical master equation~\eqref{eq:meq} by a Langevin equation for the state concentrations $x_i(t) = N_i(t)/N$ (details of the derivation are provided in Supplementary Note~1). Originally proposed for Brownian particles suspended in a liquid, the Langevin equation decomposes the dynamics of a sample trajectory of the random process into two contributions: into a deterministic drift and into noise stemming from the discreteness of particle numbers (``demographic fluctuations''). Both the demographic fluctuations and the contribution to the deterministic drift that corresponds to mutations in the EGT setting are suppressed by a small prefactor $1/N$. Therefore, these terms change the dynamics only slowly. The deterministic drift that corresponds to interactions between agents is, however, not suppressed. It thus governs the dynamics to leading order.

Hence, we find that the leading order dynamics of the condensation process~\eqref{eq:meq}-\eqref{eq:rates} are described by the differential equations:
\begin{equation}\label{eq:ALVE}
\frac{\mathrm{d}}{\mathrm{d} t} x_i = x_i (A \vec{x})_i \, .
\end{equation} 
The matrix $A$ is antisymmetric and encodes the effective transition rates between states ($a_{ij} = r_{ij}-r_{ji}$). The constants $s_{ij}$ that occur in the definition of the rates~\eqref{eq:rates} do not change the leading order dynamics, but they become relevant on subleading order timescales.

We refer to equation~\eqref{eq:ALVE} as the antisymmetric Lotka-Volterra equation (ALVE). It provides a description of pairwise interactions that preserve the total number of particles. Therefore, the ALVE finds a broad range of applications in diverse fields of research, in addition to the aforementioned condensation of bosons far from equilibrium. It was first studied by Volterra\cite{Volterra1931} in the context of predator-prey oscillations in population biology\cite{Goel1971, May, Reichenbach2006}. In plasma physics, the ALVE describes the spectra of plasma oscillations (Langmuir waves)\cite{Zakharov1974, Manakov1975}, and in chemical kinetics it captures the dynamics of bimolecular autocatalytic reactions\cite{Itoh1971, DiCera1988,DiCera1989,VanKampen2007}. In EGT, the ALVE is known as the replicator equation of zero-sum games such as the rock-paper-scissors game\cite{Akin1984,Chawanya2002, Reichenbach2006, Knebel2013}. Table~\ref{table:1} summarizes all of the above analogies. 

Despite the simple structure of the ALVE, it exhibits a rich and complex behaviour. In the following, we show how the mathematical analysis of the ALVE explains condensation into multiple states (condensate selection). To this end, we extend an approach for the analysis of the ALVE that was introduced in the context of EGT\cite{Akin1984,Chawanya2002}. 

\subsection{Production of relative entropy and condensate selection.}\
Our analysis starts from a theorem in linear programming theory\cite{Tucker1956}.
Given an antisymmetric matrix $A$, it is always possible to find a vector $\vec{c}$ that fulfils the following conditions: its entries are positive for indices in $I\subseteq\{1,\dots, S\}$  and zero for indices in $\bar{I}=\{1, \dots, S\}-I$, whereas the entries of $A\vec{c}$ are zero for indices in $I$ and negative for indices in $\bar{I}$ (Fig.~\ref{fig:1}b). Although several vectors $\vec{c}$ with these properties may exist, the index set $I$ is unique and, thus, determined by the antisymmetric matrix $A$.
Finding such a ``condensate vector'' $\vec{c}$ is crucial for the understanding of condensate selection and of the condensation dynamics. The condensate vector has the following physical interpretations.

All condensate vectors yield fixed points of the ALVE~\eqref{eq:ALVE}. Because of the antisymmetry of matrix $A$, a linear stability analysis of these fixed points does not yield insight into the global dynamics (Supplementary Note~1). However, global stability properties can be inferred by showing that the relative entropy of a condensate vector to the state concentrations,
\begin{equation}\label{eq:D}
D(\vec{c}||\vec{x}) = \sum_{i\in I} c_i\log{\left(c_i / x_i\right)} \, ,
\end{equation}
is a Lyapunov function (note that we do not consider the relative entropy of the state concentrations to the condensate vector, but define the relative entropy vice versa). 
The relative entropy~\eqref{eq:D} decreases with time and is bounded from below (see Methods and Supplementary Fig.~1).  Therefore, the dynamics relax to a subsystem in which relative entropy production is zero. The relaxation of relative entropy production is reminiscent of Prigogine's study of open systems in non-equilibrium thermodynamics. Indeed, we find that the system, to cite Prigogine's phrase, ``settles down to the state of least dissipation''\cite{Prigogine1978}. 

This state of least dissipation is characterized even further by the condensate vector $\vec{c}$. 
Considering the definition of the relative entropy~\eqref{eq:D} and its boundedness, it follows that every concentration $x_i$ with $i\in I$ remains larger than a positive constant.
On the other hand, states with indices in $\bar{I}$ become depleted for long times (see Methods).
Therefore, we find that the condensate vector determines the selection of condensates. Positive entries of $\vec{c}$ correspond to states that become condensates, whereas zero entries of $\vec{c}$ correspond to states that become depleted. Both the set of condensates and the set of depleted states are unique (Fig.~\ref{fig:1}b) and independent of the initial conditions. Generically, the entries of the condensate vector are also unique upon normalization (its entries sum up to one) and yield the rate $|(A\vec{c})_i|$ at which a state $E_i$ becomes depleted. The condensate selection occurs exponentially fast (see Fig.~\ref{fig:2} and Methods). 

After relaxation, the dynamics of the system are restricted to the condensates. In other words, the condensates form the attractor of the dynamics.
However, the dynamics in this subsystem do not come to rest. The state of least dissipation is a dynamic state with a perpetually changing number of particles in the condensates: periodic, quasiperiodic, and non-periodic oscillations are observed (Fig.~\ref{fig:2}b and Supplementary Fig.~1). In the generic case, the entries of the condensate vector represent the temporal average of condensate concentrations according to the ALVE~\eqref{eq:ALVE}.
After condensate selection, the dynamics of these active condensates take place on a high-dimensional, deformed sphere\cite{Knebel2013}.

\subsection{An algebraic algorithm to find the condensates.}\ 
Numerical integration of the ALVE~\eqref{eq:ALVE} is neither a feasible nor a reliable method for identifying condensates (Fig.~\ref{fig:2}, Supplementary Figs.~1 and~2). Instead, we determine these states by numerically searching for a condensate vector $\vec{c}$. To this end, we reformulate the above conditions on $\vec{c}$ in terms of two linear inequalities\cite{Tucker1956}: 
\begin{equation}\label{eq:tucker}
A\vec{c} \leq 0 \text{~~and~~} \vec{c} - A\vec{c} > 0 \ .
\end{equation} 
We solve these inequalities with a linear programming algorithm that is both reliable and efficient. The time to find a condensate vector scales only polynomially with the number of states~$S$ (see Supplementary Fig.~3 and Methods for details).

\subsection{Condensation in large random networks of states.}\ 
We used our combined analytical and numerical approach to study how the connectivity of a random network of states affects the selection of condensates under the dynamics of the ALVE~\eqref{eq:ALVE}. The connectivity specifies the percentage of states between which particle transitions occur\cite{Gardner1970, Albert2002}. After having generated a network with a given connectivity, the strength and direction of an allowed transition between states $E_j$ and $E_i$ were determined by randomly sampling the corresponding effective rate constant $a_{ij}=r_{ij}-r_{ji}$. 

Our results for condensation in large random networks of states are summarized in Fig.~\ref{fig:3}. When the connectivity of a network is zero, all of its states are isolated. Particles are not exchanged between states and none of the states becomes depleted.
For an increased connectivity, isolated pairs of states are sampled in a random network. One state in an isolated pair is always depleted and the average number of condensates decreases rapidly.
Upon approaching a critical connectivity, cycles and trees of all orders become embedded in a random network. This critical connectivity scales inversely with the number of states\cite{Albert2002}.
We observe that, under the dynamics of the ALVE~\eqref{eq:ALVE}, the average number of condensates becomes minimal for a connectivity that also scales inversely with the number of states (see Fig.~\ref{fig:3}g). 
We attribute this minimum to the interplay between the criticality of random networks and condensate selection on connected components of the network\cite{Durney2011, Knebel2013}.
Embedded directed cycles are a recurring motif\cite{Milo2002} in the remaining network of condensates after condensate selection.
Above the critical connectivity, a single giant cluster is formed. On average, half the number of states in this giant cluster become condensates once the network is fully connected ($C=1$) (Fig.~\ref{fig:3}a)\cite{Chawanya2002, Allesina2011, Vorberg2013}. Thus, our analysis emphasizes the importance of critical properties of random networks for condensate selection.

\subsection{Design of active condensates.}
Our understanding of condensate selection can be used to design systems that condense into a particular network of states, a game of condensates. We exemplify this procedure by formulating conditions under which a system relaxes into a rock-paper-scissors (RPS) game of condensates\cite{Reichenbach2007}. Three particular states $E_1, E_2,$ and $E_3$ in a system become a RPS cycle of condensates if, and only if, the following two conditions are fulfilled (Fig.~\ref{fig:4}). First, the ``RPS condition'' requires that the rate constants between the three states form a RPS network (for example, $r_{12}>r_{21},r_{23}>r_{32},$ and $r_{31}>r_{13}$). Second, the ``attractivity condition'' requires that the inflow of particles into the RPS cycle from any other state $E_k$ is greater than the outflow to that state $E_k$ (for all $k=4, \dots, S$). The values of the rate constants between the states that become depleted are irrelevant. More complex games of condensates can be designed by formulating similar conditions on the rate constants. These conditions are formulated as inequalities that depend on Pfaffians of the antisymmetric matrix $A$ and its submatrices (see Methods)\cite{Knebel2013}. The flow of particles between states in these systems causes condensate concentrations to oscillate~(Fig.~\ref{fig:2}b).

\section*{Discussion}

Our findings thus suggest intriguing dynamics of condensates in systems whose temporal evolutions are captured by the classical master equation~\eqref{eq:meq} with rates~\eqref{eq:rates}; for example in driven-dissipative systems of non-interacting bosons. 
Condensates observed on the leading order timescale are metastable. For longer times, relaxation into a steady state occurs\cite{Vorberg2013, Eisert2015}. When detailed balance is broken in the system of condensates, the net probability current between at least two states does not vanish and a non-equilibrium steady state is approached\cite{Zia2007, Kriecherbauer2010}. 
The simplest way of designing such condensates is illustrated by the above RPS game. In this game, detailed balance is broken, for example, when the transition of particles is unidirectional (with totally asymmetric rate constants $r_{12}>r_{21}=0,r_{23}>r_{32}=0,$ and $r_{31}>r_{13}=0$).  
For non-interacting bosons in driven-dissipative systems, the continuing supply with energy through the external time-periodic driving force (Floquet system) and the dissipation of energy into the environment  may, therefore, prevent the system from reaching equilibrium. How such systems may be realized in an experiment poses an interesting question for future research.

The transition of particles between condensates in the here studied coupled birth-death processes parallels the interaction and mutation of winning agents in evolutionary game theory, reflecting an ``evolutionary game of condensates''. Our results suggest the possibility of creating novel bosonic systems with an oscillating occupation of condensates. Non-interacting bosons in incoherently driven-dissipative systems are promising candidates. Since the antisymmetric Lotka-Volterra equation also arises in population biology, chemical kinetics, and plasma physics, all of our mathematical results apply to these fields as well. 

\begin{methods}
%
\subsection{Asymptotics of the antisymmetric Lotka-Volterra equation.}
The asymptotic behaviour of the ALVE~\eqref{eq:ALVE} can be characterized as follows: For every antisymmetric matrix $A$ there exists a unique subset of states $I\subseteq\{1, \dots, S\}$ whose concentrations stay away from zero for all times, that is,
\begin{align}
x_i(t) \geq Const(A, \vec{x}_0) >0  \text{ for all } t\geq 0 \text{ and for every } i\in I\ .
\end{align}
The set $I$ is the set of condensates. All of the other states with indices in $\bar{I}=\{1, \dots, S\}-I$ become depleted as $t \rightarrow \infty$, that is, 
\begin{align}
 x_i(t)\to 0  \text{ as } t\to \infty \text{ for every } i\in \bar{I}\ .
\end{align}
The set of condensates can be determined algebraically from the antisymmetric matrix $A$ and does not depend on the initial conditions $\vec{x}_0\in \Delta_{S-1}=\{\vec{x}\in \mathbb{R}^S\ |\ x_i>0 \text{ for all }i,\ \sum_{i=1}^S x_i=1\}$.

To show this result, the time-dependent entropy $D(\vec{c}||\vec{x})(t)$ of a condensate vector $\vec{c} = (c_1, \dots, c_n) \in\overline{\Delta}_{S-1} $ ($c_i\geq 0$ for all $i$ and $\sum_i c_i =1$) relative to the trajectory $\vec{x}(t)$ is considered (that is, the Kullback-Leibler divergence of $\vec{x}(t)$ from $\vec{c}$), see equation~\eqref{eq:D}.
A condensate vector is defined via the properties (see Fig.~\ref{fig:1}b):
\begin{align}\label{eq:b_1}
c_i&>0\ \text{ and } (A\vec{c})_i=0  \text{ for } i\in I \text{, and} \\\label{eq:b_2}
c_i&=0\ \text{ and }  (A\vec{c})_i<0 \text{ for } i\in \bar{I}\  .
\end{align}
Such a vector can always be found for an antisymmetric matrix\cite{Tucker1956}. Notably, the index set $I$ is unique although more than one condensate vector may exist.

Considering the time derivative of the relative entropy $D(\vec{c}||\vec{x})(t)$ and employing equations~\eqref{eq:ALVE} and \eqref{eq:b_1} yields: 
\begin{align}\label{eq:D_derivative}
\frac{\dif}{\dif t}D(\vec{c}||\vec{x})(t) &= -\sum_{i=1}^S c_i\frac{\partial_t x_i}{ x_i}= -\sum_{i=1}^S c_i (A\vec{x})_i = \sum_{i=1}^S (A\vec{c})_i x_i = \sum_{i\in \bar{I}}  (A\vec{c})_i x_i\ .
\end{align} 
Since $(A\vec{c})_{\bar{I}}<0$ and $\vec{x}>0$, it follows that $\partial_t D(\vec{c}||\vec{x})(t)< 0$. Therefore, the relative entropy $D(\vec{c}||\vec{x})$ is a Lyapunov function if $\vec{c}$ is chosen in accordance with equations~\eqref{eq:b_1} and~\eqref{eq:b_2}. Moreover, $D(\vec{c}||\vec{x})$ is bounded from above by $D(\vec{c}||\vec{x})(0)$ and from below by zero for all times. This can be seen from the definition of $D$, and  from integration of equation~\eqref{eq:D_derivative} (using that $ (A\vec{c})_{\bar{I}}<0$ and $\vec{x}>0$):
\begin{align}\label{eq:D_boundedness}
0\leq D(\vec{c}||\vec{x})(t) = D(\vec{c}||\vec{x})(0) +\int_0^t ds\ \sum_{i\in \bar{I}}  (A\vec{c})_i x_i(s)\leq D(\vec{c}||\vec{x})(0)\ .
\end{align} 

From the definition of the relative entropy in equation~\eqref{eq:D}, it follows that every concentration $x_i$ with $i\in I$ remains larger than a positive constant, that is, $x_i(t)\geq Const(A, \vec{x}_0) >0$ for all times~$t$ (if $x_i(t)\to 0$ for $i\in I$, it follows that $D\to \infty$, which contradicts the boundedness of $D$).

Furthermore, equation~\eqref{eq:D_boundedness} implies that,
\begin{align}
-\int_0^t ds\ (A\vec{c})_i x_i(s)
 \leq -\int_0^t ds\ \sum_{i\in \bar{I}}  (A\vec{c})_i x_i(s)
 \leq D(\vec{c}||\vec{x})(0)\ ,
\end{align} 
for every $i\in \bar{I}$  and for all $t$.
%
%
Therefore, concentration $x_i$ is integrable for every $i\in \bar{I}$ ($x_i\in L^1(0,\infty)$) with the bound:
\begin{align}\label{eq:x_int}
0<\int_0^\infty ds\  x_i(s) \leq \frac{D(\vec{c}||\vec{x})(0)}{-(A\vec{c})_i} = Const(A, \vec{x}_0) \quad \text{for every } i\in \bar{I}\ .
\end{align} 
Since the derivative of the concentrations is bounded from above and below,  $|\partial_t x_i| = |x_i (A\vec{x})_i| \leq \lVert(A\vec{x}) \rVert_\infty \leq  \lVert A\rVert_{\infty\to\infty}\leq Const(A)$, one concludes that $x_i$ is uniformly continuous ($\lVert A\rVert_{\infty\to\infty}$ denotes the operator norm of $A$ induced by the maximum norm on $\mathbb{R}^S$). Together with the integrability~\eqref{eq:x_int}, it follows that states with indices in $\bar{I}$ become depleted as $t \rightarrow \infty$, that is, $x_i(t)\to 0$ for $i\in \bar{I}$.

In conclusion, given an antisymmetric matrix $A=R-R^T$ via a rate constant matrix $R=\{r_{ij}\}_{i,j}$, one finds a condensate vector $\vec{c}$ that satisfies inequalities~\eqref{eq:b_1}-\eqref{eq:b_2}. The index set $I$, for which entries of $\vec{c}$ are positive, represents condensates. The index set $\bar{I}$, for which entries of $\vec{c}$ are zero, represents states that become depleted. Moreover, equation~\eqref{eq:D_derivative} implies that the relative entropy becomes a conserved quantity in the subsystem of condensates (Supplementary Fig.~1).

\subsection{Temporal average of condensate concentrations.}

The ALVE~\eqref{eq:ALVE} is solved implicitly by,
\begin{align}\label{eq:RE_integrated}
x_i(t) = x_i(0) e^{ t\cdot  (A\langle\vec{x}\rangle_t)_i}\ ,
\end{align}
with the time average of the trajectory $\langle\vec{x}\rangle_t$ defined as: 
\begin{align}
\langle\vec{x}\rangle_t = \frac{1}{t}\int_0^t ds\ \vec{x}(s)\ .
\end{align}

It is shown above that $0<Const(A, \vec{x}_0)\leq x_i(t)\leq1$ holds for the states that become condensates ($i \in I$). By rearranging equation~\eqref{eq:RE_integrated}, one thus obtains:
\begin{align}\label{eq:long_time_average1}
|(A\langle\vec{x}\rangle_t)_i|\leq \frac{1}{t}\left|\, \log{\left(\frac{x_i(t)}{x_i(0)}\right)}\right|\leq \frac{Const(A, \vec{x}_0)}{t} \quad \text{for all } i\in I\ .
\end{align}
Note that $Const$ is used to denote arbitrary positive, time-independent constants. Therefore, the right hand side of equation~\eqref{eq:long_time_average1} vanishes  for $t\to \infty$. 
On the other hand, $x_i$ is integrable for $i \in \bar{I}$ (equation~\eqref{eq:x_int}). Thus, the corresponding component of the time average converges to zero,
\begin{align}\label{eq:long_time_average2}
\langle x_i \rangle_t \leq \frac{Const(A, \vec{x}_0)}{t}\to 0  \text{ as } t\to \infty \text{ for every } i\in \bar{I}\ .
\end{align}
Hence, the distance of the time average $\langle\vec{x}\rangle_t$ to the kernel of the antisymmetric submatrix $A^I$  converges to zero (the submatrix $A^I$ corresponds to the system of condensates with indices in $I$).

\subsection{Structure of a generic antisymmetric matrix.}
For systems with an even number of states $S$, the antisymmetric matrix $A = R-R^T$ generically has a trivial kernel, whereas for systems with an odd number of states, the kernel of $A$ is generically one-dimensional. A higher dimensional kernel of $A$ only occurs if the matrix entries are tuned\cite{Goel1971, Knebel2013, Cullis1913}. 
As a consequence, when all of the entries above the diagonal of $A$ are, for example, randomly drawn from a continuous probability distribution (for example from a Gaussian distribution), all $2^S$ submatrices of $A$ have a kernel with dimension of less than or equal to one.

The projection of $\vec{x}\in \mathbb{R}^S$ to the subspace $\mathbb{R}^{(J)}\subseteq\mathbb{R}^S$ for an arbitrary index set $J\subseteq\{1, \dots, S\}$ is defined as $\vec{x}_J \coloneqq P_J\vec{x} \coloneqq (x_j)_{j\in J}$. In other words, entries of $\vec{x}_J$ are zero for indices in the complement~$\bar{J}$. In the following, the short notation $A^J \coloneqq P_J A P_J$ is also used (see above). Furthermore, the set of antisymmetric matrices whose submatrices have a kernel with dimension less than or equal to one is defined: 
\begin{align}\label{eq:omega}
\Omega \coloneqq \left\{ A \in \mathbb{R}^{S\times S} \ |\ A \text{ is antisymmetric and } \dim \ker  A^J\leq 1  \text{ for all } J\subseteq \{1,\dots , S\}  \right\} 
 \ .
 \end{align}
The complement $\bar{\Omega}$ has measure zero with respect to the flat measure $\dif A$ on antisymmetric matrices (the translation invariant measure, which is  sigma-finite and not trivial).

For antisymmetric matrices $A\in\Omega$, the kernel can be characterized as follows\cite{Knebel2013, Cullis1913}. 
If the number of states $S$ is even, the kernel of $A$ is trivial: $\ker{A} = \{\vec{0}\}$. If the number of states is odd, the kernel is one-dimensional: $\ker{A} = \{\vec{v}\}$. This kernel element can be computed analytically in terms of Pfaffians of submatrices of $A$:
\begin{align}\label{eq:kernel}
\vec{v} = \left(\Pf{A_{\hat{1}}}, -\Pf{A_{\hat{2}}}, \dots, \Pf{A_{\hat{S}}}\right)\ .
 \end{align}
The submatrix $A_{\hat{k}}\in\mathbb{R}^{(S-1)\times (S-1)}$ denotes the matrix for which the $k$-th column and row are removed from $A$. 

For antisymmetric matrices $A\in\Omega$, the normalized condensate vector $\vec{c}$ with $\sum_i c_i = 1$ and with properties \eqref{eq:b_1}, \eqref{eq:b_2} is unique. The latter follows from $A^I\vec{c}=0$ (equations~\eqref{eq:b_1} and~\eqref{eq:omega}).  Therefore, the condensate vector is the unique kernel vector of the subsystem of condensates whose interactions are characterized by the matrix $A^I$. Furthermore, $I$ contains an odd number of elements.
To determine the condensate vector for $A\in\Omega$, one can proceed as follows. For each odd-dimensional submatrix $A^I$  with $I\subseteq \{1,\dots , S\}$, one computes the kernel element $\vec{v}$ according to equation~\eqref{eq:kernel} and defines the vector $\vec{w}\in\mathbb{R}^{S}$ by setting $\vec{w}_I=\vec{v}$ and $\vec{w}_{\bar{I}}=0$. There exists exactly one set $I$ for which $(A\vec{w})_{\bar{I}}<0$. The corresponding vector $\vec{w}$ is the unique condensate vector upon normalization.\\

\subsection{Temporal average of condensate concentrations (generic case).}
It was shown above that the temporal average of condensate concentrations $\langle\vec{x}\rangle_t$ converges to a non-negative kernel element of the antisymmetric matrix $A^I$. In the generic case, the condensate vector $\vec{c}$ is the unique kernel element of $A^I$ upon normalization. Therefore, positive entries of $\vec{c}$ represent the asymptotic temporal average of condensate concentrations, 
\begin{align}\label{eq:kernel_estimate}
\lVert \mean{\vec{x}}_t - \vec{c}\rVert_{\infty}\leq \frac{Const(A, \vec{x}_0)}{t} \to 0 \text{ as } t\to\infty\ .
\end{align}
%


\subsection{Exponentially fast depletion of states (generic case).}

Upon inserting equation~\eqref{eq:kernel_estimate} into the implicit solution~\eqref{eq:RE_integrated} of the ALVE, the exponentially fast depletion of states with $i\in \bar{I}$ can be seen as follows (note that $(A\vec{c})_i < 0$ according to the choice of the condensate vector in equations~\eqref{eq:b_1} and~\eqref{eq:b_2}):
\begin{align}
x_i(t) &= x_i(0) e^{ t\cdot  (A\langle\vec{x}\rangle_t)_i}\\
&\leq x_i(0) e^{ t\cdot(  (A\vec{c})_i + \lVert A(\langle \vec{x} \rangle_t-\vec{c})\rVert_{\infty})}\\
& \leq x_i(0) e^{ t\cdot  (A\vec{c})_i + Const(A, \vec{x}_0)}\\
& = Const(A, \vec{x}_0) e^{t\cdot  (A\vec{c})_i}\ , 
\end{align}
and analogously,
\begin{align}
x_i(t) &= x_i(0) e^{ t\cdot  (A\langle\vec{x}\rangle_t)_i}\geq  Const(A, \vec{x}_0) e^{t\cdot  (A\vec{c})_i}\ .
\end{align}
Therefore, condensate selection occurs exponentially fast at depletion rate $|(A\vec{c})_i|$. 
The dynamics of cases for non-generic antisymmetric matrices are discussed in Supplementary Note~2.\\

\subsection{Linear programming algorithm.}
For the numeric computation of condensate vectors $\vec{c}$, a finite threshold $\delta > 0$ was introduced into the inequalities~\eqref{eq:tucker}: $A\vec{c} \leq 0$ and $\vec{c} - A\vec{c}  \geq \delta > 0 $. Its value was set to $\delta = 1$ by rescaling of $\vec{c}$. Numerical solution of the inequalities was performed by using the IBM ILOG CPLEX Optimization Studio 12.5 and its interface to the C++ language. The software Mathematica 9.0 from Wolfram Research was also found to be applicable. Further information on the calibration of the linear programming algorithm and a simplified Mathematica algorithm are provided in Supplementary Note~3.\\

\end{methods}


\clearpage
\section*{References}


\begin{thebibliography}{10}
\expandafter\ifx\csname url\endcsname\relax
  \def\url#1{\texttt{#1}}\fi
\expandafter\ifx\csname urlprefix\endcsname\relax\def\urlprefix{URL }\fi
\providecommand{\bibinfo}[2]{#2}
\providecommand{\eprint}[2][]{\url{#2}}

\bibitem{Krapivsky2000}
\bibinfo{author}{Krapivsky, P.~L.}, \bibinfo{author}{Redner, S.} \&
  \bibinfo{author}{Leyvraz, F.}
\newblock \bibinfo{title}{Connectivity of growing random networks}.
\newblock \emph{\bibinfo{journal}{Phys. Rev. Lett.}}
  \textbf{\bibinfo{volume}{85}}, \bibinfo{pages}{4629--4632}
\newblock  (\bibinfo{year}{2000}).

\bibitem{Bianconi2001}
\bibinfo{author}{Bianconi, G.} \& \bibinfo{author}{Barab\'asi, A.-L.}
\newblock \bibinfo{title}{Bose-{E}instein condensation in complex networks}.
\newblock \emph{\bibinfo{journal}{Phys. Rev. Lett.}}
  \textbf{\bibinfo{volume}{86}}, \bibinfo{pages}{5632--5635}
\newblock  (\bibinfo{year}{2001}).

\bibitem{Evans2005}
\bibinfo{author}{Evans, M.~R.} \& \bibinfo{author}{Hanney, T.}
\newblock \bibinfo{title}{Nonequilibrium statistical mechanics of the
  zero-range process and related models}.
\newblock \emph{\bibinfo{journal}{J. Phys. A: Math. Gen.}}
  \textbf{\bibinfo{volume}{38}}, \bibinfo{pages}{R195--R240}
\newblock  (\bibinfo{year}{2005}).

\bibitem{Evans1996}
\bibinfo{author}{Evans, M.~R.}
\newblock \bibinfo{title}{Bose-{E}instein condensation in disordered exclusion
  models and relation to traffic flow}.
\newblock \emph{\bibinfo{journal}{Europhys. Lett.}}
  \textbf{\bibinfo{volume}{36}}, \bibinfo{pages}{13--18}
\newblock  (\bibinfo{year}{1996}).

\bibitem{Krug1996}
\bibinfo{author}{Krug, J.} \& \bibinfo{author}{Ferrari, P.~A.}
\newblock \bibinfo{title}{Phase transitions in driven diffusive systems with
  random rates}.
\newblock \emph{\bibinfo{journal}{J. Phys. A: Math. Gen.}}
  \textbf{\bibinfo{volume}{29}}, \bibinfo{pages}{L465--L471}
\newblock  (\bibinfo{year}{1996}).

\bibitem{Chowdhury2000}
\bibinfo{author}{Chowdhury, D.}, \bibinfo{author}{Santen, L.} \&
  \bibinfo{author}{Schadschneider, A.}
\newblock \bibinfo{title}{Statistical physics of vehicular traffic and some
  related systems}.
\newblock \emph{\bibinfo{journal}{Phys. Rep.}} \textbf{\bibinfo{volume}{329}},
  \bibinfo{pages}{199--329}
\newblock  (\bibinfo{year}{2000}).

\bibitem{Kaupuzs2005}
\bibinfo{author}{Kaupu\ifmmode~\check{z}\else \v{z}\fi{}s, J.},
  \bibinfo{author}{Mahnke, R.} \& \bibinfo{author}{Harris, R.~J.}
\newblock \bibinfo{title}{Zero-range model of traffic flow}.
\newblock \emph{\bibinfo{journal}{Phys. Rev. E}} \textbf{\bibinfo{volume}{72}},
  \bibinfo{pages}{056125}
\newblock  (\bibinfo{year}{2005}).

\bibitem{Spitzer1970}
\bibinfo{author}{Spitzer, F.}
\newblock \bibinfo{title}{Interaction of {M}arkov processes}.
\newblock \emph{\bibinfo{journal}{Advances in Mathematics}}
  \textbf{\bibinfo{volume}{5}}, \bibinfo{pages}{246--290}
\newblock  (\bibinfo{year}{1970}).

\bibitem{Evans2014}
\bibinfo{author}{Evans, M.~R.} \& \bibinfo{author}{Waclaw, B.}
\newblock \bibinfo{title}{Condensation in stochastic mass transport models:
  beyond the zero-range process}.
\newblock \emph{\bibinfo{journal}{J. Phys. A: Math. Theor.}}
  \textbf{\bibinfo{volume}{47}}, \bibinfo{pages}{095001}
\newblock  (\bibinfo{year}{2014}).

\bibitem{Bose1924}
\bibinfo{author}{Bose, S.~N.}
\newblock \bibinfo{title}{Plancks {G}esetz und {L}ichtquantenhypothese}.
\newblock \emph{\bibinfo{journal}{Z. Phys.}} \textbf{\bibinfo{volume}{26}},
  \bibinfo{pages}{178--181}
\newblock  (\bibinfo{year}{1924}).

\bibitem{Einstein1924}
\bibinfo{author}{Einstein, A.}
\newblock \bibinfo{title}{{Quantentheorie des einatomigen idealen Gases}}.
\newblock \emph{\bibinfo{journal}{Sitzb. d. Preuss. Akad. d. Wiss.}}
  \bibinfo{pages}{261--267}
\newblock  (\bibinfo{year}{1924}).

\bibitem{Einstein1925}
\bibinfo{author}{Einstein, A.}
\newblock \bibinfo{title}{{Quantentheorie des einatomigen idealen Gases. Zweite
  Abhandlung}}.
\newblock \emph{\bibinfo{journal}{Sitzb. d. Preuss. Akad. d. Wiss.}}
  \bibinfo{pages}{3--14}
\newblock  (\bibinfo{year}{1925}).

\bibitem{Griffin1995}
\bibinfo{author}{Griffin, A.}, \bibinfo{author}{Snoke, D.} \&
  \bibinfo{author}{Stringari, G.}
\newblock \emph{\bibinfo{title}{{Bose Einstein Condensation}}}
  (\bibinfo{publisher}{Cambridge University Press},
  \bibinfo{address}{Cambridge}, \bibinfo{year}{1995}).

\bibitem{Anglin2002}
\bibinfo{author}{Anglin, J.~R.} \& \bibinfo{author}{Ketterle, W.}
\newblock \bibinfo{title}{Bose-{E}instein condensation of atomic gases}.
\newblock \emph{\bibinfo{journal}{Nature}} \textbf{\bibinfo{volume}{416}},
  \bibinfo{pages}{211--218}
\newblock  (\bibinfo{year}{2002}).

\bibitem{Penrose1956}
\bibinfo{author}{Penrose, O.} \& \bibinfo{author}{Onsager, L.}
\newblock \bibinfo{title}{Bose-{E}instein condensation and liquid helium}.
\newblock \emph{\bibinfo{journal}{Phys. Rev.}} \textbf{\bibinfo{volume}{104}},
  \bibinfo{pages}{576--584}
\newblock  (\bibinfo{year}{1956}).

\bibitem{Mueller2006}
\bibinfo{author}{Mueller, E.~J.}, \bibinfo{author}{Ho, T.-L.},
  \bibinfo{author}{Ueda, M.} \& \bibinfo{author}{Baym, G.}
\newblock \bibinfo{title}{Fragmentation of {B}ose-{E}instein condensates}.
\newblock \emph{\bibinfo{journal}{Phys. Rev. A}} \textbf{\bibinfo{volume}{74}},
  \bibinfo{pages}{033612}
\newblock  (\bibinfo{year}{2006}).

\bibitem{Gardiner}
\bibinfo{author}{Gardiner, C.}
\newblock \emph{\bibinfo{title}{{Stochastic Methods: A Handbook for the Natural
  and Social Sciences}}} (\bibinfo{publisher}{Springer},
  \bibinfo{address}{Berlin}, \bibinfo{year}{2009}).

\bibitem{VanKampen2007}
\bibinfo{author}{{Van Kampen}, N.~G.}
\newblock \emph{\bibinfo{title}{{Stochastic Processes in Physics and Chemistry}}}
  (\bibinfo{publisher}{Elsevier}, \bibinfo{address}{Amsterdam},
  \bibinfo{year}{2007}).

\bibitem{Vorberg2013}
\bibinfo{author}{Vorberg, D.}, \bibinfo{author}{Wustmann, W.},
  \bibinfo{author}{Ketzmerick, R.} \& \bibinfo{author}{Eckardt, A.}
\newblock \bibinfo{title}{Generalized {B}ose-{E}instein condensation into
  multiple states in driven-dissipative systems}.
\newblock \emph{\bibinfo{journal}{Phys. Rev. Lett.}}
  \textbf{\bibinfo{volume}{111}}, \bibinfo{pages}{240405}
\newblock  (\bibinfo{year}{2013}).

\bibitem{Albert2002}
\bibinfo{author}{Albert, R.} \& \bibinfo{author}{Barab\'asi, A.-L.}
\newblock \bibinfo{title}{Statistical mechanics of complex networks}.
\newblock \emph{\bibinfo{journal}{Rev. Mod. Phys.}}
  \textbf{\bibinfo{volume}{74}}, \bibinfo{pages}{47--97}
\newblock  (\bibinfo{year}{2002}).

\bibitem{Milo2002}
\bibinfo{author}{Milo, R.} \emph{et~al.}
\newblock \bibinfo{title}{Network motifs: simple building blocks of complex
  networks}.
\newblock \emph{\bibinfo{journal}{Science}} \textbf{\bibinfo{volume}{298}},
  \bibinfo{pages}{824--827}
\newblock  (\bibinfo{year}{2002}).

\bibitem{Bluemel1991}
\bibinfo{author}{Bl\"umel, R.} \emph{et~al.}
\newblock \bibinfo{title}{Dynamical localization in the microwave interaction
  of {R}ydberg atoms: the influence of noise}.
\newblock \emph{\bibinfo{journal}{Phys. Rev. A}} \textbf{\bibinfo{volume}{44}},
  \bibinfo{pages}{4521--4540}
\newblock  (\bibinfo{year}{1991}).

\bibitem{Kohler1997}
\bibinfo{author}{Kohler, S.}, \bibinfo{author}{Dittrich, T.} \&
  \bibinfo{author}{H\"anggi, P.}
\newblock \bibinfo{title}{Floquet-{M}arkovian description of the parametrically
  driven, dissipative harmonic quantum oscillator}.
\newblock \emph{\bibinfo{journal}{Phys. Rev. E}} \textbf{\bibinfo{volume}{55}},
  \bibinfo{pages}{300--313}
\newblock  (\bibinfo{year}{1997}).

\bibitem{Breuer2000}
\bibinfo{author}{Breuer, H.-P.}, \bibinfo{author}{Huber, W.} \&
  \bibinfo{author}{Petruccione, F.}
\newblock \bibinfo{title}{Quasistationary distributions of dissipative
  nonlinear quantum oscillators in strong periodic driving fields}.
\newblock \emph{\bibinfo{journal}{Phys. Rev. E}} \textbf{\bibinfo{volume}{61}},
  \bibinfo{pages}{4883--4889}
\newblock  (\bibinfo{year}{2000}).

\bibitem{Grifoni1998}
\bibinfo{author}{Grifoni, M.} \& \bibinfo{author}{H{\"a}nggi, P.}
\newblock \bibinfo{title}{Driven quantum tunneling}.
\newblock \emph{\bibinfo{journal}{Phys. Rep.}} \textbf{\bibinfo{volume}{304}},
  \bibinfo{pages}{229--354}
\newblock  (\bibinfo{year}{1998}).

\bibitem{Breuer2006}
\bibinfo{author}{Breuer, H.-P.} \& \bibinfo{author}{Petruccione, F.}
\newblock \emph{\bibinfo{title}{{The Theory of Open Quantum Systems}}}
  (\bibinfo{publisher}{Oxford University Press}, \bibinfo{address}{Oxford},
  \bibinfo{year}{2002}).

\bibitem{Gardiner1997}
\bibinfo{author}{Gardiner, C.~W.} \& \bibinfo{author}{Zoller, P.}
\newblock \bibinfo{title}{Quantum kinetic theory: a quantum kinetic master
  equation for condensation of a weakly interacting {B}ose gas without a
  trapping potential}.
\newblock \emph{\bibinfo{journal}{Phys. Rev. A}} \textbf{\bibinfo{volume}{55}},
  \bibinfo{pages}{2902--2921}
\newblock  (\bibinfo{year}{1997}).

\bibitem{Kagan1997}
\bibinfo{author}{Kagan, Y.} \& \bibinfo{author}{Svistunov, B.~V.}
\newblock \bibinfo{title}{Evolution of correlation properties and appearance of
  broken symmetry in the process of {B}ose-{E}instein condensation}.
\newblock \emph{\bibinfo{journal}{Phys. Rev. Lett.}}
  \textbf{\bibinfo{volume}{79}}, \bibinfo{pages}{3331--3334}
\newblock  (\bibinfo{year}{1997}).

\bibitem{Bijlsma2000}
\bibinfo{author}{Bijlsma, M.~J.}, \bibinfo{author}{Zaremba, E.} \&
  \bibinfo{author}{Stoof, H. T.~C.}
\newblock \bibinfo{title}{Condensate growth in trapped {B}ose gases}.
\newblock \emph{\bibinfo{journal}{Phys. Rev. A}} \textbf{\bibinfo{volume}{62}},
  \bibinfo{pages}{063609}
\newblock  (\bibinfo{year}{2000}).

\bibitem{Gardiner1998}
\bibinfo{author}{Gardiner, C.~W.}, \bibinfo{author}{Lee, M.~D.},
  \bibinfo{author}{Ballagh, R.~J.}, \bibinfo{author}{Davis, M.~J.} \&
  \bibinfo{author}{Zoller, P.}
\newblock \bibinfo{title}{Quantum kinetic theory of condensate growth:
  comparison of experiment and theory}.
\newblock \emph{\bibinfo{journal}{Phys. Rev. Lett.}}
  \textbf{\bibinfo{volume}{81}}, \bibinfo{pages}{5266--5269}
\newblock  (\bibinfo{year}{1998}).

\bibitem{Walser1999}
\bibinfo{author}{Walser, R.}, \bibinfo{author}{Williams, J.},
  \bibinfo{author}{Cooper, J.} \& \bibinfo{author}{Holland, M.}
\newblock \bibinfo{title}{Quantum kinetic theory for a condensed bosonic gas}.
\newblock \emph{\bibinfo{journal}{Phys. Rev. A}} \textbf{\bibinfo{volume}{59}},
  \bibinfo{pages}{3878--3889}
\newblock  (\bibinfo{year}{1999}).

\bibitem{Kocharovsky2000}
\bibinfo{author}{Kocharovsky, V.~V.}, \bibinfo{author}{Scully, M.~O.},
  \bibinfo{author}{Zhu, S.-Y.} \& \bibinfo{author}{Suhail~Zubairy, M.}
\newblock \bibinfo{title}{Condensation of {N} bosons. {II}. {N}onequilibrium
  analysis of an ideal {B}ose gas and the laser phase-transition analogy}.
\newblock \emph{\bibinfo{journal}{Phys. Rev. A}} \textbf{\bibinfo{volume}{61}},
  \bibinfo{pages}{023609}
\newblock  (\bibinfo{year}{2000}).

\bibitem{Pauli1928}
\bibinfo{author}{Pauli, W.}
\newblock \emph{\bibinfo{title}{{Festschrift zum 60. Geburtstage A.
  Sommerfeld}}} (\bibinfo{address}{Hirzel, Leipzig}, \bibinfo{year}{1928}).

\bibitem{Mandel1995}
\bibinfo{author}{Mandel, L.} \& \bibinfo{author}{Wolf, E.}
\newblock \emph{\bibinfo{title}{Optical Coherence and Quantum Optics}}
  (\bibinfo{publisher}{Cambridge University Press},
  \bibinfo{address}{Cambridge, UK}, \bibinfo{year}{1995}).

\bibitem{Gardiner2004}
\bibinfo{author}{Gardiner, C.~W.} \& \bibinfo{author}{Zoller, P.}
\newblock \emph{\bibinfo{title}{{Quantum Noise}}}
  (\bibinfo{publisher}{Springer}, \bibinfo{address}{Berlin Heidelberg},
  \bibinfo{year}{2004}).

\bibitem{Smith1982}
\bibinfo{author}{Maynard~Smith, J.}
\newblock \emph{\bibinfo{title}{{Evolution and the Theory of Games}}}
  (\bibinfo{publisher}{Cambridge University Press},
  \bibinfo{address}{Cambridge, UK}, \bibinfo{year}{1982}).

\bibitem{Nowak2004}
\bibinfo{author}{Nowak, M.~A.} \& \bibinfo{author}{Sigmund, K.}
\newblock \bibinfo{title}{Evolutionary dynamics of biological games}.
\newblock \emph{\bibinfo{journal}{Science}} \textbf{\bibinfo{volume}{303}},
  \bibinfo{pages}{793--799}
\newblock  (\bibinfo{year}{2004}).

\bibitem{Sinervo1996}
\bibinfo{author}{Sinervo, B.} \& \bibinfo{author}{Lively, C.~M.}
\newblock \bibinfo{title}{The rock-paper-scissors game and the evolution of
  alternative male strategies}.
\newblock \emph{\bibinfo{journal}{Nature}} \textbf{\bibinfo{volume}{380}},
  \bibinfo{pages}{240--243}
\newblock  (\bibinfo{year}{1996}).

\bibitem{Kerr2002}
\bibinfo{author}{Kerr, B.}, \bibinfo{author}{Riley, M.},
  \bibinfo{author}{Feldman, M.} \& \bibinfo{author}{Bohannan, B.}
\newblock \bibinfo{title}{Local dispersal promotes biodiversity in a real-life
  game of rock-paper-scissors}.
\newblock \emph{\bibinfo{journal}{Nature}} \textbf{\bibinfo{volume}{418}},
  \bibinfo{pages}{171--174}
\newblock  (\bibinfo{year}{2002}).

\bibitem{Reichenbach2007}
\bibinfo{author}{Reichenbach, T.}, \bibinfo{author}{Mobilia, M.} \&
  \bibinfo{author}{Frey, E.}
\newblock \bibinfo{title}{Mobility promotes and jeopardizes biodiversity in
  rock-paper-scissors games}.
\newblock \emph{\bibinfo{journal}{Nature}} \textbf{\bibinfo{volume}{448}},
  \bibinfo{pages}{1046--1049}
\newblock  (\bibinfo{year}{2007}).

\bibitem{Weber2014}
\bibinfo{author}{Weber, M.~F.}, \bibinfo{author}{Poxleitner, G.},
  \bibinfo{author}{Hebisch, E.}, \bibinfo{author}{Frey, E.} \&
  \bibinfo{author}{Opitz, M.}
\newblock \bibinfo{title}{Chemical warfare and survival strategies in bacterial
  range expansions}.
\newblock \emph{\bibinfo{journal}{J. R. Soc. Interface}}
  \textbf{\bibinfo{volume}{11}}, \bibinfo{pages}{20140172}
\newblock  (\bibinfo{year}{2014}).

\bibitem{Szolnoki2014}
\bibinfo{author}{Szolnoki, A.} \emph{et~al.}
\newblock \bibinfo{title}{Cyclic dominance in evolutionary games: a review}.
\newblock \emph{\bibinfo{journal}{J. R. Soc. Interface}}
  \textbf{\bibinfo{volume}{11}}, \bibinfo{pages}{20140735}
\newblock  (\bibinfo{year}{2014}).

\bibitem{Nowak2004paper}
\bibinfo{author}{Nowak, M.~A.}, \bibinfo{author}{Sasaki, A.},
  \bibinfo{author}{Taylor, C.} \& \bibinfo{author}{Fudenberg, D.}
\newblock \bibinfo{title}{Emergence of cooperation and evolutionary stability
  in finite populations}.
\newblock \emph{\bibinfo{journal}{Nature}} \textbf{\bibinfo{volume}{428}},
  \bibinfo{pages}{646--650}
\newblock  (\bibinfo{year}{2004}).

\bibitem{Szolnoki2014b}
\bibinfo{author}{Szolnoki, A.}, \bibinfo{author}{Antonioni, A.},
  \bibinfo{author}{Tomassini, M.} \& \bibinfo{author}{Perc, M.}
\newblock \bibinfo{title}{Binary birth-death dynamics and the expansion of
  cooperation by means of self-organized growth}.
\newblock \emph{\bibinfo{journal}{EPL}} \textbf{\bibinfo{volume}{105}},
  \bibinfo{pages}{48001}
\newblock  (\bibinfo{year}{2014}).

\bibitem{McKane2005}
\bibinfo{author}{McKane, A.~J.} \& \bibinfo{author}{Newman, T.~J.}
\newblock \bibinfo{title}{Predator-prey cycles from resonant amplification of
  demographic stochasticity}.
\newblock \emph{\bibinfo{journal}{Phys. Rev. Lett.}}
  \textbf{\bibinfo{volume}{94}}, \bibinfo{pages}{218102}
\newblock  (\bibinfo{year}{2005}).

\bibitem{Traulsen2005}
\bibinfo{author}{Traulsen, A.}, \bibinfo{author}{Claussen, J.~C.} \&
  \bibinfo{author}{Hauert, C.}
\newblock \bibinfo{title}{Coevolutionary dynamics: from finite to infinite
  populations}.
\newblock \emph{\bibinfo{journal}{Phys. Rev. Lett.}}
  \textbf{\bibinfo{volume}{95}}, \bibinfo{pages}{238701}
\newblock  (\bibinfo{year}{2005}).

\bibitem{Reichenbach2006}
\bibinfo{author}{Reichenbach, T.}, \bibinfo{author}{Mobilia, M.} \&
  \bibinfo{author}{Frey, E.}
\newblock \bibinfo{title}{Coexistence versus extinction in the stochastic
  cyclic {L}otka-{V}olterra model}.
\newblock \emph{\bibinfo{journal}{Phys. Rev. E}} \textbf{\bibinfo{volume}{74}},
  \bibinfo{pages}{51907}
\newblock  (\bibinfo{year}{2006}).

\bibitem{Melbinger2010}
\bibinfo{author}{Melbinger, A.}, \bibinfo{author}{Cremer, J.} \&
  \bibinfo{author}{Frey, E.}
\newblock \bibinfo{title}{Evolutionary game theory in growing populations}.
\newblock \emph{\bibinfo{journal}{Phys. Rev. Lett.}}
  \textbf{\bibinfo{volume}{105}}, \bibinfo{pages}{178101}
\newblock  (\bibinfo{year}{2010}).

\bibitem{Biancalani2014}
\bibinfo{author}{Biancalani, T.}, \bibinfo{author}{Dyson, L.} \&
  \bibinfo{author}{McKane, A.~J.}
\newblock \bibinfo{title}{Noise-induced bistable states and their mean
  switching time in foraging colonies}.
\newblock \emph{\bibinfo{journal}{Phys. Rev. Lett.}}
  \textbf{\bibinfo{volume}{112}}, \bibinfo{pages}{038101}
\newblock  (\bibinfo{year}{2014}).

\bibitem{Rulands2014}
\bibinfo{author}{Rulands, S.}, \bibinfo{author}{Jahn, D.} \&
  \bibinfo{author}{Frey, E.}
\newblock \bibinfo{title}{Specialization and bet hedging in heterogeneous
  populations}.
\newblock \emph{\bibinfo{journal}{Phys. Rev. Lett.}}
  \textbf{\bibinfo{volume}{113}}, \bibinfo{pages}{108102}
\newblock  (\bibinfo{year}{2014}).

\bibitem{Volterra1931}
\bibinfo{author}{Volterra, V.}
\newblock \emph{\bibinfo{title}{{Le{\c c}ons sur la Th{\'e}orie
  Math{\'e}matique de la Lutte pour la Vie}}}
  (\bibinfo{publisher}{Gauthier-Villars}, \bibinfo{address}{Paris},
  \bibinfo{year}{1931}).

\bibitem{Goel1971}
\bibinfo{author}{Goel, N.~S.}, \bibinfo{author}{Maitra, S.~C.} \&
  \bibinfo{author}{Montroll, E.~W.}
\newblock \bibinfo{title}{On the {V}olterra and other nonlinear models of
  interacting populations}.
\newblock \emph{\bibinfo{journal}{Rev. Mod. Phys.}}
  \textbf{\bibinfo{volume}{43}}, \bibinfo{pages}{231--276}
\newblock  (\bibinfo{year}{1971}).

\bibitem{May}
\bibinfo{author}{May, R.~M.}
\newblock \emph{\bibinfo{title}{{Stability and Complexity in Model
  Ecosystems}}} (\bibinfo{publisher}{Princeton University Press},
  \bibinfo{address}{Princeton, NJ}, \bibinfo{year}{1973}).

\bibitem{Zakharov1974}
\bibinfo{author}{Zakharov, V.}, \bibinfo{author}{Musher, S.} \&
  \bibinfo{author}{Rubenchik, A.}
\newblock \bibinfo{title}{Nonlinear stage of parametric wave excitation in a
  plasma}.
\newblock \emph{\bibinfo{journal}{JETP Lett.}} \textbf{\bibinfo{volume}{19}},
  \bibinfo{pages}{151--152} (\bibinfo{year}{1974}).

\bibitem{Manakov1975}
\bibinfo{author}{Manakov, S.}
\newblock \bibinfo{title}{Complete integrability and stochastization of
  discrete dynamical systems}.
\newblock \emph{\bibinfo{journal}{Sov. Phys.-JETP}}
  \textbf{\bibinfo{volume}{40}}, \bibinfo{pages}{269--274}
  (\bibinfo{year}{1975}).

\bibitem{Itoh1971}
\bibinfo{author}{Itoh, Y.}
\newblock \bibinfo{title}{Boltzmann equation on some algebraic structure
  concerning struggle for existence}.
\newblock \emph{\bibinfo{journal}{Proc. Japan Acad.}}
  \textbf{\bibinfo{volume}{47}}, \bibinfo{pages}{854--858}
\newblock  (\bibinfo{year}{1971}).

\bibitem{DiCera1988}
\bibinfo{author}{Di~Cera, E.}, \bibinfo{author}{Phillipson, P.~E.} \&
  \bibinfo{author}{Wyman, J.}
\newblock \bibinfo{title}{Chemical oscillations in closed macromolecular
  systems}.
\newblock \emph{\bibinfo{journal}{Proc. Natl. Acad. Sci. USA}}
  \textbf{\bibinfo{volume}{85}}, \bibinfo{pages}{5923--5926}
  (\bibinfo{year}{1988}).

\bibitem{DiCera1989}
\bibinfo{author}{Di~Cera, E.}, \bibinfo{author}{Phillipson, P.~E.} \&
  \bibinfo{author}{Wyman, J.}
\newblock \bibinfo{title}{Limit-cycle oscillations and chaos in reaction
  networks subject to conservation of mass}.
\newblock \emph{\bibinfo{journal}{Proc. Natl. Acad. Sci. USA}}
  \textbf{\bibinfo{volume}{86}}, \bibinfo{pages}{142--146}
  (\bibinfo{year}{1989}).

\bibitem{Akin1984}
\bibinfo{author}{Akin, E.} \& \bibinfo{author}{Losert, V.}
\newblock \bibinfo{title}{Evolutionary dynamics of zero-sum games}.
\newblock \emph{\bibinfo{journal}{J. Math. Biol.}}
  \textbf{\bibinfo{volume}{20}}, \bibinfo{pages}{231--258}
\newblock  (\bibinfo{year}{1984}).

\bibitem{Chawanya2002}
\bibinfo{author}{Chawanya, T.} \& \bibinfo{author}{Tokita, K.}
\newblock \bibinfo{title}{Large-dimensional replicator equations with
  antisymmetric random interactions}.
\newblock \emph{\bibinfo{journal}{J. Phys. Soc. Jpn.}}
  \textbf{\bibinfo{volume}{71}}, \bibinfo{pages}{429--431}
\newblock  (\bibinfo{year}{2002}).

\bibitem{Knebel2013}
\bibinfo{author}{Knebel, J.}, \bibinfo{author}{Kr\"uger, T.},
  \bibinfo{author}{Weber, M.~F.} \& \bibinfo{author}{Frey, E.}
\newblock \bibinfo{title}{Coexistence and survival in conservative
  {L}otka-{V}olterra networks}.
\newblock \emph{\bibinfo{journal}{Phys. Rev. Lett.}}
  \textbf{\bibinfo{volume}{110}}, \bibinfo{pages}{168106}
\newblock  (\bibinfo{year}{2013}).

\bibitem{Tucker1956}
\bibinfo{author}{Kuhn, H.} \& \bibinfo{author}{Tucker, A.}
\newblock \emph{\bibinfo{title}{{Linear Inequalities and Related Systems}}}
  (\bibinfo{publisher}{Princeton University Press},
  \bibinfo{address}{Princeton, NJ}, \bibinfo{year}{1956}).

\bibitem{Prigogine1978}
\bibinfo{author}{Prigogine, I.}
\newblock \bibinfo{title}{Time, structure, and fluctuations}.
\newblock \emph{\bibinfo{journal}{Science}} \textbf{\bibinfo{volume}{201}},
  \bibinfo{pages}{777--785}
\newblock  (\bibinfo{year}{1978}).

\bibitem{Gardner1970}
\bibinfo{author}{Gardner, M.~R.} \& \bibinfo{author}{Ashby, W.~R.}
\newblock \bibinfo{title}{Connectance of large dynamic (cybernetic) systems:
  critical values for stability}.
\newblock \emph{\bibinfo{journal}{Nature}} \textbf{\bibinfo{volume}{228}},
  \bibinfo{pages}{784}
\newblock  (\bibinfo{year}{1970}).

\bibitem{Durney2011}
\bibinfo{author}{Durney, C.~H.}, \bibinfo{author}{Case, S.~O.},
  \bibinfo{author}{Pleimling, M.} \& \bibinfo{author}{Zia, R. K.~P.}
\newblock \bibinfo{title}{Saddles, arrows, and spirals: deterministic
  trajectories in cyclic competition of four species}.
\newblock \emph{\bibinfo{journal}{Phys. Rev. E}} \textbf{\bibinfo{volume}{83}},
  \bibinfo{pages}{051108}
\newblock  (\bibinfo{year}{2011}).

\bibitem{Allesina2011}
\bibinfo{author}{Allesina, S.} \& \bibinfo{author}{Levine, J.~M.}
\newblock \bibinfo{title}{A competitive network theory of species diversity}.
\newblock \emph{\bibinfo{journal}{Proc. Natl. Acad. Sci. USA}}
  \textbf{\bibinfo{volume}{108}}, \bibinfo{pages}{5638--5642}
\newblock  (\bibinfo{year}{2011}).

\bibitem{Eisert2015}
\bibinfo{author}{Eisert, J.}, \bibinfo{author}{Friesdorf, M.} \&
  \bibinfo{author}{Gogolin, C.}
\newblock \bibinfo{title}{Quantum many-body systems out of equilibrium}.
\newblock \emph{\bibinfo{journal}{Nature Phys.}} \textbf{\bibinfo{volume}{11}},
  \bibinfo{pages}{124--130}
\newblock  (\bibinfo{year}{2015}).

\bibitem{Zia2007}
\bibinfo{author}{Zia, R. K.~P.} \& \bibinfo{author}{Schmittmann, B.}
\newblock \bibinfo{title}{Probability currents as principal characteristics in
  the statistical mechanics of non-equilibrium steady states}.
\newblock \emph{\bibinfo{journal}{J. Stat. Mech.}}
  \textbf{\bibinfo{volume}{2007}}, \bibinfo{pages}{P07012}
\newblock  (\bibinfo{year}{2007}).

\bibitem{Kriecherbauer2010}
\bibinfo{author}{Kriecherbauer, T.} \& \bibinfo{author}{Krug, J.}
\newblock \bibinfo{title}{A pedestrian's view on interacting particle systems,
  {KPZ} universality and random matrices}.
\newblock \emph{\bibinfo{journal}{J. Phys. A: Math. Theor.}}
  \textbf{\bibinfo{volume}{43}}, \bibinfo{pages}{403001}
\newblock  (\bibinfo{year}{2010}).

\bibitem{Cullis1913}
\bibinfo{author}{{Cullis}, C.~E.}
\newblock \emph{\bibinfo{title}{{Matrices and Determinoids}}}, vol.
  \bibinfo{volume}{I and II} (\bibinfo{publisher}{Cambridge University Press},
  \bibinfo{address}{Cambridge}, \bibinfo{year}{1913}).

\end{thebibliography}

\providecommand{\noopsort}[1]{}\providecommand{\singleletter}[1]{#1}%

\clearpage
\begin{addendum}
\item[Supplementary Information] accompanies this paper at http://www.nature.com/naturecommunications
%
 \item[Acknowledgements] 
We are grateful for fruitful discussions with Peter Zoller, Ulrich Schollw\"ock, Immanuel Bloch, Wilhelm Zwerger, Andr\'e Eckardt, Daniel Vorberg, Alexander Schnell, Marianne Bauer, Brendan Osberg, Jacob Halatek, Matthias Bauer, and Sebastian Koch. 
This work was supported by the Deutsche Forschungsgemeinschaftas as project A7 of the SFB TR 12 ``Symmetry and Universality in Mesoscopic Systems'', and by the German Excellence Initiative via the program ``Nanosystems Initiative Munich'' (NIM). J.K. acknowledges funding from the Studienstiftung des Deutschen Volkes.
%
\item[Author Contributions] 
J.K., M.F.W., T.K., E.F. designed, discussed, and planned the study. T.K., J.K., M.F.W. developed the analytical results. M.F.W., T.K., J.K. developed the numerical algorithms and generated the data. J.K., M.F.W., T.K., E.F. interpreted the results and wrote the manuscript.
%
\item[Reprints and permission] 
information is available online at http://npg.nature.com/reprintsandpermissions/\\ 
The authors declare no competing financial interests.\\\\ Correspondence and requests for materials
should be addressed to E.F.~(email: frey@lmu.de).
\end{addendum}


\clearpage

\begin{figure}
\centerline{\includegraphics{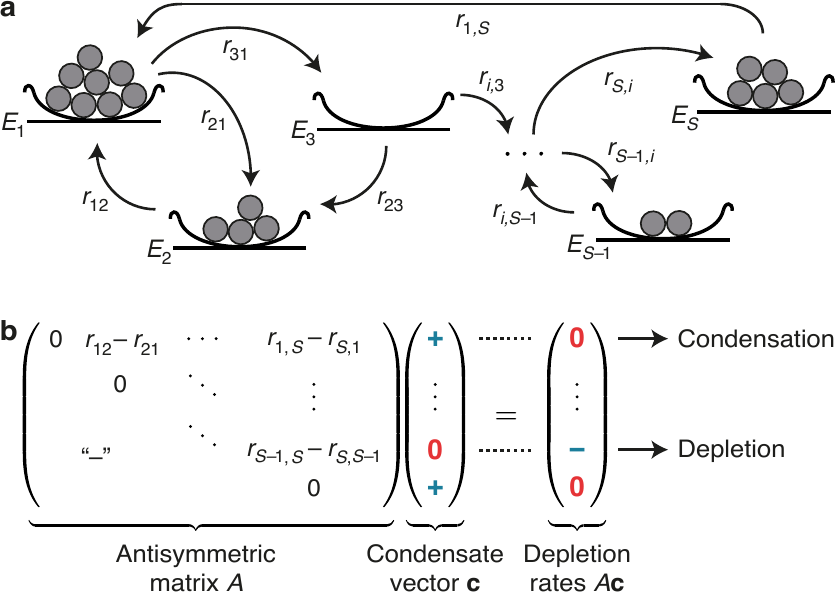}}
\end{figure}
%
%
%
\noindent {\bf Figure 1. Condensation into multiple states due to particle transitions between states and mathematics of condensate selection.} 
(\textbf{a}) 
With respect to condensation in an incoherently driven-dissipative quantum system, each bowl represents a state $E_i$ that is occupied by $N_i$ non-interacting bosons (filled circles). If indicated by an arrow, bosons may undergo transitions from state $E_j$ to state $E_i$ at a rate $\Gamma_{i \leftarrow j} = r_{ij} (N_i+1) N_j$, with rate constant $r_{ij}$. 
In the language of evolutionary game theory, the figure depicts the interaction of $N_i$ agents (filled circles) playing strategies $E_i$ (bowls). An agent playing strategy $E_j$ adopts strategy $E_i$ at a rate $\Gamma_{i \leftarrow j} = r_{ij} N_i N_j$. The above rate of bosonic condensate selection is recovered if agents may also spontaneously mutate from $E_j$ to $E_i$ at a rate $r_{ij}$.
(\textbf{b}) 
A condensate vector $\vec{c}$ for an antisymmetric matrix $A$ has two properties: its entries are positive for indices for which $A\vec{c}$ is zero, and they are zero for indices for which $A\vec{c}$ is negative (``-''~signifies the antisymmetry of matrix $A$). Temporal evolution of the relative entropy of the condensate vector to the state concentrations under the ALVE~\eqref{eq:ALVE} relates positive entries of the condensate vector to condensates, and its zero entries to depleted states. Generically, positive entries of $\vec{c}$ represent the asymptotic temporal average of oscillating condensate concentrations  according to the ALVE~\eqref{eq:ALVE}, and negative entries of $A\vec{c}$ represent depletion rates.

\clearpage
%
%
\begin{figure}
\centerline{\includegraphics{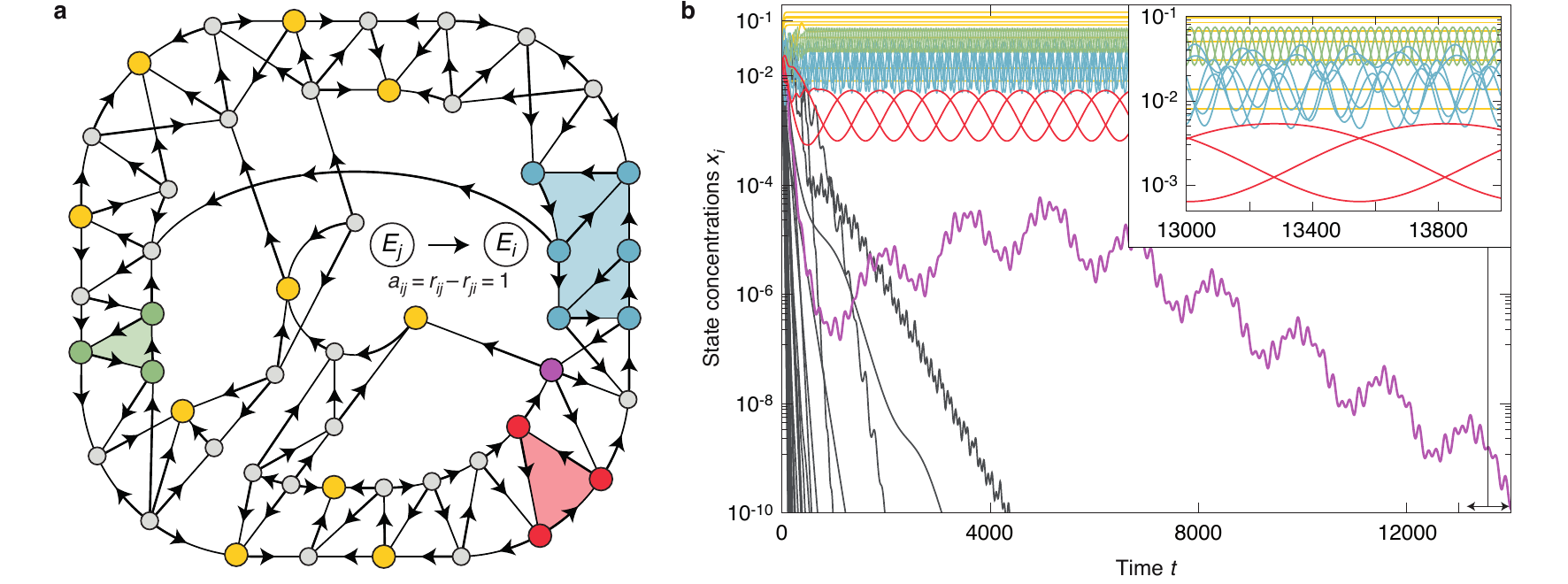}}
\end{figure}
%
%
%
\noindent {\bf Figure 2. Fragmentation of an exemplary system into multiple condensates with oscillating state concentrations.} 
(\textbf{a}) Randomly sampled network of $50$ states.  Disks represent states. An arrow from state $E_j$ to state $E_i$ represents an effective rate constant $a_{ij} = r_{ij}-r_{ji} = 1$ (a missing arrow indicates a forbidden transition with $a_{ij}=0$). Computation of a condensate vector~$\vec{c}$ predicted relaxation into ten isolated condensates (yellow), one interacting subsystem with six condensates (blue), and two rock-paper-scissors (RPS) cycles (red and green). All other states become depleted. The complete network also comprises RPS cycles of which some states become depleted. Knowledge of the network topology alone is thus insufficient to determine condensates.
(\textbf{b}) Temporal evolution of state concentrations $x_i$ (logarithmic scale). Colours in accordance with (\textbf{a}). Numerical integration of the ALVE~\eqref{eq:ALVE} confirmed the selection of states based on the condensate vector $\vec{c}$. 
Subsystems with six (blue) and three (red and green) condensates exhibit oscillations of concentrations with non-vanishing particle flow. Depletion of states occurs exponentially fast. Identifying condensates from a condensate vector $\vec{c}$ is more reliable than through numerical integration: The concentration of the state associated to the purple disk in (\textbf{a}) decays exponentially to a concentration of $1.5 \cdot 10^{-7}$ before recovering transiently. Numerical integration cannot rule out permanent recovery at later times. Supplementary Fig.~2 demonstrates such a case.

\clearpage
%
%
\begin{figure}
\centerline{\includegraphics{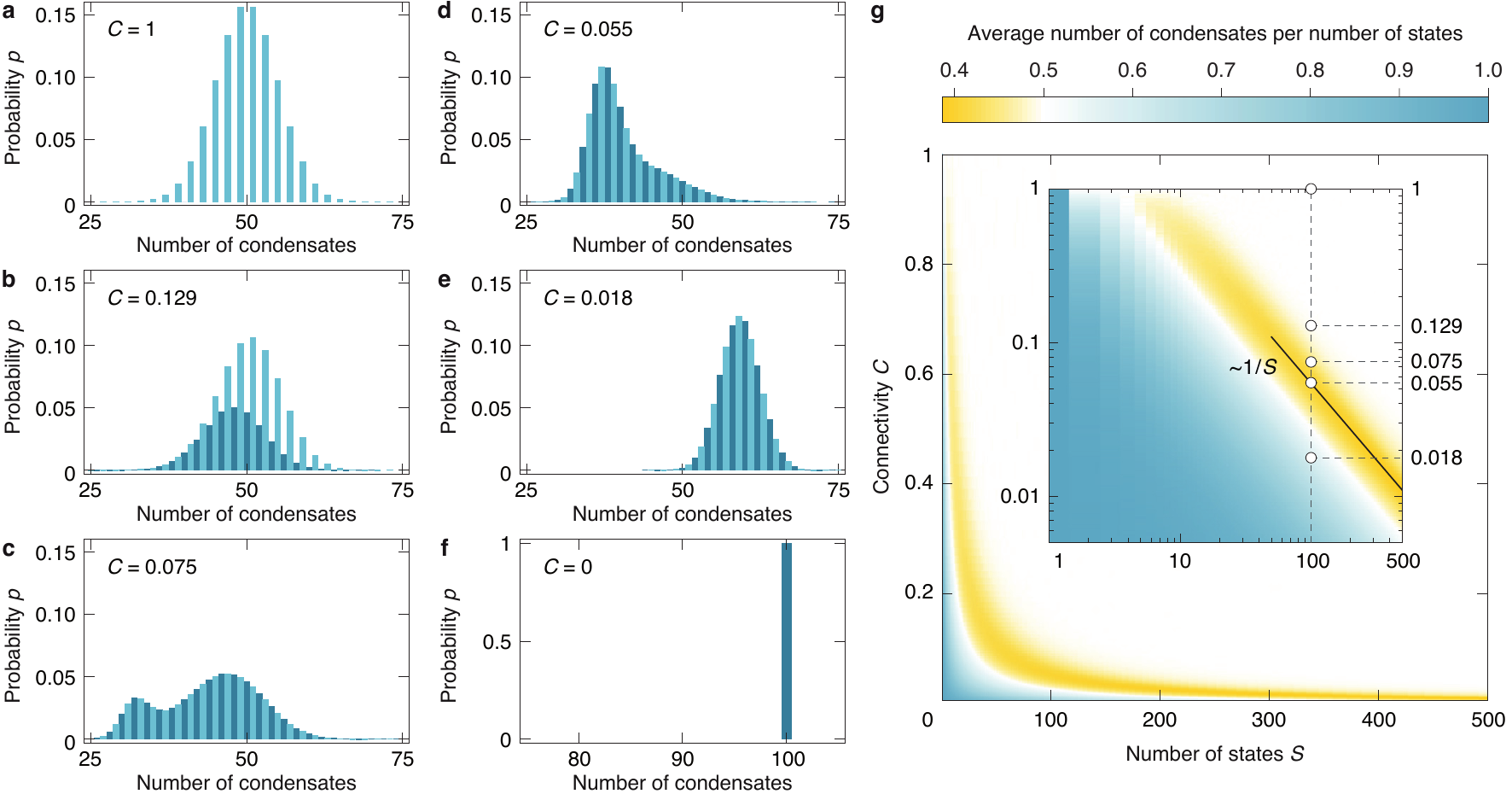}}
\end{figure}
%
%
%
\noindent {\bf Figure 3. Dependence of the number of condensates on the connectivity of states in random networks.} 
(\textbf{a-f}) Measured probability $p$ of finding a particular number of condensates for a system with $S=100$ states and with connectivity $C$ ($5\cdot 10^6$ systems analysed per histogram). The connectivity specifies the percentage of states between which transitions of particles occur with a non-zero effective rate constant $a_{ij}=r_{ij}-r_{ji}$. Effective rate constants $a_{ij}$ were sampled from a Gaussian distribution (zero mean, unit variance). (\textbf{a}) At full connectivity, the distribution is pseudo-binomial with only odd numbers of condensates ($C=1$; light blue bars)\cite{Chawanya2002, Allesina2011, Vorberg2013}. (\textbf{b}) As the connectivity is reduced, even numbers of condensates become possible when systems decouple into even numbers of subsystems ($C=0.129$; dark blue bars). (\textbf{c}) The distribution exhibits bimodality ($C=0.075$) and (\textbf{d}) approaches a minimal average number of $40.2$ condensates ($C=0.055$). (\textbf{e}) This average subsequently increases ($C=0.018$) because isolated states are trivially selected as condensates ($C=0$) as shown in (\textbf{f}). 
(\textbf{g}) Average number of condensates per number of states (colour-coded) plotted against the number of states~$S$ and the connectivity~$C$ (log-log graph in inset; $\geq10^4$ systems per data point, see Supplementary Fig.~4 for the reliability of the linear programming algorithm). White circles correspond to distributions shown in (\textbf{a-e}). The minimal relative number of condensates conforms to the power law $C \sim 1/S^\gamma$ with $\gamma=0.998\pm 0.008$ (s.e.m.) and can be related to the criticality of random networks\cite{Albert2002}.

\clearpage
%
%
\begin{figure}
\centerline{\includegraphics{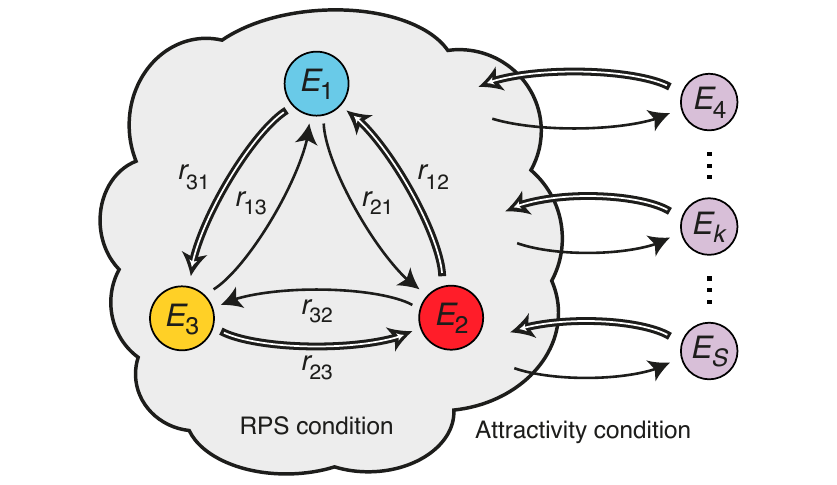}}
\end{figure}
%
%
%
\noindent {\bf Figure 4. Conditions for the emergence of a rock-paper-scissors cycle of condensates.}
Three particular states $E_1, E_2,$ and $E_3$ (blue, red, and yellow disks) of a network condense into a rock-paper-scissors (RPS) cycle if, and only if, two conditions are fulfilled: First, the ``RPS condition'' requires that the rate constants $r_{ij}$ between the three states form a RPS network: $r_{i-1,i+1}>r_{i+1,i-1}$ (indices are counted modulo 3, for example, $r_{42}=r_{12}$ (framed arrows denote rate constants that are larger than rate constants for the respective reverse direction). Differences between these rate constants define the entries $c_i=r_{i-1,i+1}-r_{i+1,i-1}$ of an admissible condensate vector $\vec{c}$. Second, the ``attractivity condition'' requires that the weighted sum of rates from any exterior state $E_k$ (purple disks) into the RPS cycle, $\sum_{j=1}^3 c_j r_{jk}$ (framed arrows), is larger than the weighted sum of outbound rates, $\sum_{j=1}^3 c_j r_{kj}$ (black arrows). In other words, the inflow of particles into the RPS cycle from any exterior state needs to be greater than the outflow to that state.

\clearpage
%
%
\begin{table}%
\noindent
\includegraphics{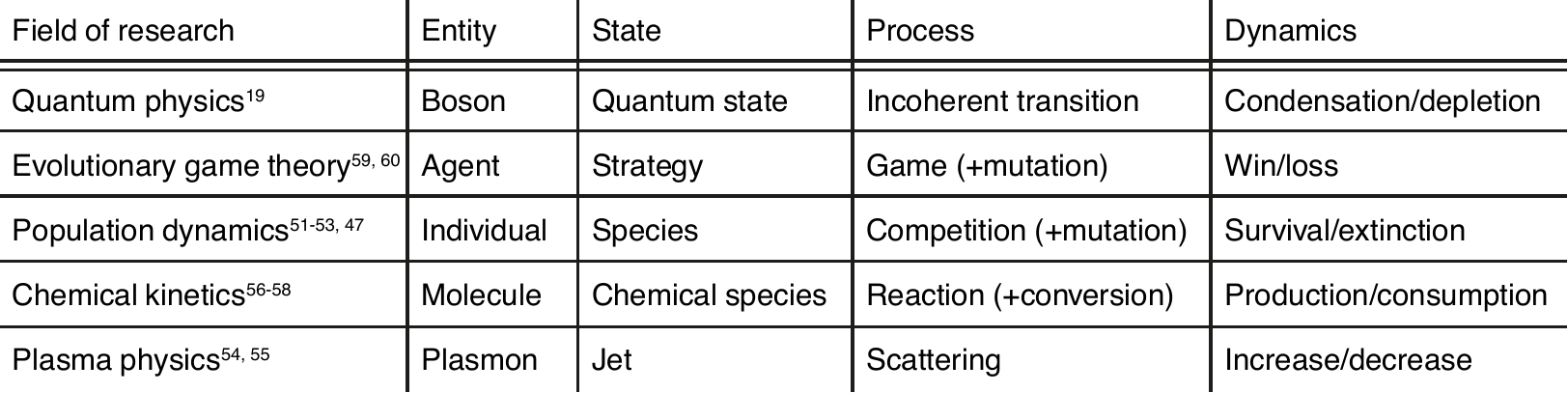}\\
\caption{\label{table:1} {\bf Condensation processes described by the ALVE in different fields of research.} The ALVE~\eqref{eq:ALVE} governs condensation processes in diverse fields of research. For example, for incoherently driven-dissipative bosonic systems, the ALVE describes condensation and depletion of states by incoherent transitions of non-interacting bosons. In EGT, the ALVE occurs in the context of winning and losing strategies played by agents. }
\end{table}

\end{document}